\documentclass{article}
\begin{document}
\title{
LUNG NODULE DETECTION IN LOW-DOSE AND HIGH-RESOLUTION CT SCANS
}
\author{
P. Delogu$^{1,2}$, M.E. Fantacci$^{1,2}$, I. Gori$^{3}$, \\
A. Preite Martinez$^{1}$, A. Retico$^{2}$ and A. Tata$^2$\\
$^{1}${\em Dipartimento di Fisica dell'Universit\`a di Pisa,} \\
{\em Largo Pontecorvo 3, 56127 Pisa, Italy}\\ 
$^{2}${\em Istituto Nazionale di Fisica Nucleare, Sezione di Pisa,} \\
{\em Largo Pontecorvo 3, 56127 Pisa, Italy}\\
$^{3}${\em Bracco Imaging S.p.A., }\\
{\em Via E. Folli 50, 20134 Milano, Italy}
}
\date{}
\maketitle
\baselineskip=11.6pt
\begin{abstract}
We are developing a computer-aided detection (CAD) system for the identification of small pulmonary nodules in screening CT scans.
The main modules of our system, i.e. a dot-enhancement filter for nodule candidate selection  and a neural classifier for false positive finding reduction, are described. The preliminary results obtained on the so-far collected database of lung CT are discussed.
\end{abstract}
\baselineskip=14pt
\section{Introduction}

Lung cancer most commonly
manifests as non-calcified pulmonary nodules. Computer Tomography (CT)
has been shown to be the best imaging modality for the detection of small pulmonary
nodules~\cite{Diederich}, particularly since the introduction of the helical technology.
The amount of images that need to be interpreted
in CT examinations can be very high, especially when multi-detector helical
CT and thin collimation are used, thus generating up to about 300 images 
per scan.
In order to support radiologists in the  identification of early-stage pathological objects, researchers have recently begun to explore computer-aided detection (CAD) methods in this area.

The First Italian Randomized Controlled Trial that aims to study the potential
impact of screening on a high-risk population using low-dose helical
CT has started last year. In this framework we are developing a CAD system for small pulmonary nodule identification. The system is based on a dot-enhancement filter and a neural-based module for the reduction of the amount of false-positive (FP) findings per 
scan.

\section{Description of the CAD system}

\begin{figure}[t]
 \vspace{6.0cm}
\includegraphics{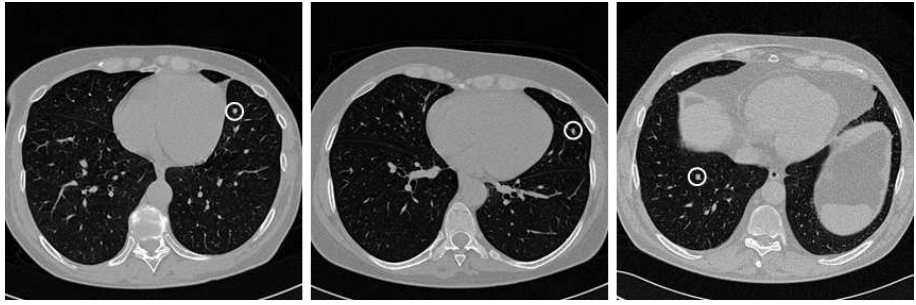}
 \caption{\it
    Some examples of small pulmonary nodules.
    \label{fig:nodules} }
\end{figure}

An important and difficult task in the automated nodule detection 
is the selection of
the nodule candidates. Nodules 
may be characterized by very low CT values and/or low contrast, 
may have CT values similar to those of
blood vessels and airway walls or may be strongly 
connected to them (see fig.~\ref{fig:nodules}).

In order to identify the nodule candidates we modeled nodules with spherical objects 
and we applied  a dot-enhancement filter to the 3D matrix of voxel data. This filter attempts to determine the local
geometrical characteristics of each voxel, by computing the eigenvalues of the
Hessian matrix and evaluating a {\it likelihood} function that was purposely
built to discriminate between local morphology of linear, planar and spherical
objects, modeled as having 3D Gaussian sections~\cite{Li}.
A simple peak-detection algorithm (i.e. a local
maxima detector) is then applied to the filter output to detect
the filtered-signal peaks.

Since most FP findings are provided by crossings between blood vessels, we attempted to reduce the amount of FP/scan by applying the procedure we  called voxel-based approach (VBA).
According to this method, each voxel of a region of interest (ROI) is characterized by the grey level intensity values of its neighborhood. 
The CT values of the voxels in a 3D neighborhood of each voxel of a ROI  are rolled down into vectors of features to be analyzed by a neural classification system. The feed-forward neural network implemented to this purpose  
 assigns each voxel either to the class of voxels belonging to a nodule, or to that of normal voxels.
We can evaluate a free response receiver operating characteristic
(FROC) curve for our CAD system, by varying the percentage of voxels
to be classified as belonging to a nodule in order to assign each ROI either to
the class of ROIs containing a nodule or normal ROIs.

\section{Results}

We tested the CAD system  on a dataset of  low-dose (screening setting: 120$\div$140 kV, 20$\div$80 mA) and high-resolution (reconstructed  slice
thickness: 1.25 mm) CT scans collected and annotated by experienced radiologists in the framework of the screening trial being conducted in Tuscany (Italy).
The database available at present for our tests consists of 20  scans, 8 of which contain  12 internal nodules. 
Each scan is a sequence of about 300 slices stored in
DICOM (Digital Imaging and COmmunications in Medicine) format.

Our preliminary results show that the 3D dot-enhancement filter 
is characterized by a high sensitivity. In particular, 
if we keep the first 67 entries of the list of interesting ROIs provided by the filter output   for each  
scan, we achieve a 100\% sensitivity to internal nodule detection.
In other words, a 100\% sensitivity is obtained at a maximum number of 67 FP/scan.

Since the amount of data available for training the neural networks in the VBA method is quite small, we first partitioned our  dataset into a train and a test set, 
then we evaluated the performances of the trained neural network both on the test set and on the whole dataset (train set + test set). 
The best results achieved are the following:  
 87.0\% sensitivity and 85.3\% specificity on the test set;
88.0\% sensitivity and 84.9\% specificity on the whole dataset.
Once the VBA approach has been applied to each ROI, 
the rate of FP findings per scan has been reduced from 67 to 14 for a sensitivity of 100\% (12 nodules detected out of 12). If the sensitivity value is decreased to 91.7\% (11 nodules detected out of 12), a rate of 8.9 FP/scan is obtained.

\section{Conclusions}

The dot-enhancement pre-processing algorithm has shown a good
sensitivity in the identification of nodule candidates, and
the VBA was shown to be an effective approach to the problem of false
positives reduction.
The results obtained so far seem promising, 
albeit they are preliminary and need to be validated
against a larger database. 
Finally, we are convinced that the methods used are effective,
and that there is margin for improvement.

\section{Acknowledgments}
We  acknowledge Dr. L. Battolla, Dr. F. Falaschi and Dr. C. Spinelli of the U.O. Radiodiagnostica 2 of the Azienda Universitaria Pisana, and Prof. D. Caramella and Dr. T. Tarantino of the Diagnostic and Interventional Radiology Division of the Dipartimento di Oncologia, Trapianti e Nuove Tecnologie in Medicina of the Pisa University.

\end{document}